\documentclass[natbib]{svjour3}                     
\smartqed  
\usepackage{graphicx}
\usepackage{color}

 \journalname{Space Science Reviews}
\begin{document}

\title{Key Atmospheric Signatures for Identifying the Source Reservoirs of Volatiles in Uranus and Neptune
}


\author{O. Mousis         	\and
        A. Aguichine        	\and
        D.~H. Atkinson 		\and
        S.~K. Atreya 		\and
        T. Cavali\'e 		\and
        J.~I. Lunine		\and
        K.~E. Mandt 		\and
        T. Ronnet 
}


\institute{O. Mousis \at
Aix Marseille Univ, CNRS, CNES, LAM, Marseille, France\\
Tel.: +33-491-055-918\\
Fax: +33-491-621-190\\
\email{olivier.mousis@lam.fr}           
\and
A. Aguichine \at
Aix Marseille Univ, CNRS, CNES, LAM, Marseille, France           
\and
D.~H. Atkinson \at
Jet Propulsion Laboratory, California Institute of Technology, 4800 Oak Grove Dr., Pasadena, CA, 91109, USA           
\and
S.~K. Atreya \at
Department of Climate and Space Sciences and Engineering, University of Michigan, Ann Arbor, MI 48109-2143, USA           
\and
T. Cavali\'e \at
Laboratoire d’Astrophysique de Bordeaux, Univ. Bordeaux, CNRS B18N, all\'ee Geoffroy Saint-Hilaire, 33615 Pessac, France\\
\at
LESIA, Observatoire de Paris, PSL Research University, CNRS, Sorbonne Universit\'es, UPMC Univ. Paris 06, Univ. Paris Diderot, Sorbonne Paris Cit\'e, F-92195 Meudon, France
\and
J.~I. Lunine \at
Department of Astronomy, Cornell University, Ithaca, NY 14853, USA
\and
K.~E. Mandt \at
Applied Physics Laboratory, Johns Hopkins University, 11100 Johns Hopkins Rd., Laurel, MD 20723, USA
\and
T. Ronnet  \at
Lund Observatory, Department of Astronomy and Theoretical Physics, Lund University, Box 43, 221 00 Lund, Sweden}

\date{Received: date / Accepted: date}

\maketitle

\begin{abstract}
We investigate the enrichment patterns of several delivery scenarios of the volatiles to the atmospheres of ice giants, having in mind that the only well constrained determination made remotely, i.e. the carbon abundance measurement, suggests that their envelopes possess highly supersolar metallicities, i.e. close to two orders of magnitude above that of the protosolar nebula. In the framework of the core accretion model, only the delivery of volatiles in solid forms (amorphous ice, clathrates, pure condensates) to these planets can account for the apparent supersolar metallicity of their envelopes. In contrast, because of the inward drift of icy particles through various snowlines, all mechanisms invoking the delivery of volatiles in vapor forms predict subsolar abundances in the envelopes of Uranus and Neptune. Alternatively, even if the disk instability mechanism remains questionable in our solar system, it might be consistent with the supersolar metallicities observed in Uranus and Neptune, assuming the two planets suffered subsequent erosion of their H-He envelopes. The enrichment patterns derived for each delivery scenario considered should be useful to interpret future in situ measurements by atmospheric entry probes. 

\end{abstract}

\section{Introduction}

The ice giant planets Uranus and Neptune represent a largely unexplored class of planetary objects that bridges the gap between the larger gas giants and the smaller terrestrial worlds. Uranus and Neptune's great heliocentric distances have made exploration challenging, being limited to flybys by the Voyager 2 mission in 1986 and 1989, respectively \citep{Ty86,Li87,Sm86,Sm89,Li92,St86,St89}. Hence, most of our knowledge of atmospheric processes taking place on these distant planets is derived from remote sensing measurements acquired from Earth-based observatories and space telescopes \citep{En00,Ka09,Ka11,Fl10,Fl14,Fe13,Or14a,Or14b,Sr14,Lel10,Le15,Ir18,Ir19a,Ir19b}. As a consequence, our knowledge of their bulk composition is dramatically limited, resulting in a very crude understanding of their formation conditions and evolution \citep{At20}.

To improve this situation, ground-truth measurements carried out in these distant planets by an atmospheric probe are needed, similar to the Galileo probe at Jupiter. Remote observations cannot provide direct, unambiguous measurements of the vertical atmospheric structure (temperatures and winds), bulk composition and cloud properties. The well-mixed atmosphere for measuring the bulk composition, hence the elemental abundances, is much deeper than the deepest levels ($\sim$1 bar) reached by remote sensing measurements \citep{At20}. With the exception of CH$_4$ \citep{Ka09,Ka11,Sr11}, and to some extent H$_2$S \citep{Ir18,Ir19a,Ir19b}, remote observations have never been able to detect the key volatile species (NH$_3$, H$_2$O) thought to compose the deep ice giant clouds. Also, noble gases are out of reach of remote sensing measurements. This is an important issue because these species provide extremely important insights concerning the formation and evolution conditions of the solar system bodies \citep{Pe91,Ow99,Ga01,Ma17,Ma15,Mo09a,Mo19} (see also \cite{Ma20}, this issue). In the meantime, both NASA and ESA agencies are expressing their interest in sending of a joint flagship mission in the 2030s that would include an atmospheric entry probe as an element of a larger orbiter to be dropped toward the ice giants (see \cite{Si20}, this issue). In this context, the aim of this paper is to review the different delivery scenarios of the volatiles to the ice giants and to derive the corresponding fingerprints in their atmospheres. The measurements of such fingerprints by atmospheric entry probes would pose important constraints on the formation scenarios of Uranus and Neptune.

Section \ref{abund} is dedicated to a brief summary of the known atmospheric abundances in both Uranus and Neptune. In Section \ref{scenar}, we review the different scenarios of volatiles delivery proposed in the literature and that can be applied to the ice giants. We also investigate further a mechanism of radial drift/diffusion for the main volatiles of interest (H$_2$O, CO, N$_2$, H$_2$S, Ar, Kr, and Xe) around their respective snowlines in the protosolar nebula (PSN). This mechanism has already been used to explain the distribution of volatiles throughout the protosolar nebula \citep{St88,Cy98,Cy99,Al14,Mo19}. Section \ref{Feat} is devoted to the inference of the atmospheric fingerprints for each of the considered scenarios of volatiles delivery. Section \ref{Disc} is dedicated to discussion and conclusions.

\section{Atmospheric Abundances}
\label{abund}

\begin{figure}
\center
\includegraphics[width=0.9\textwidth]{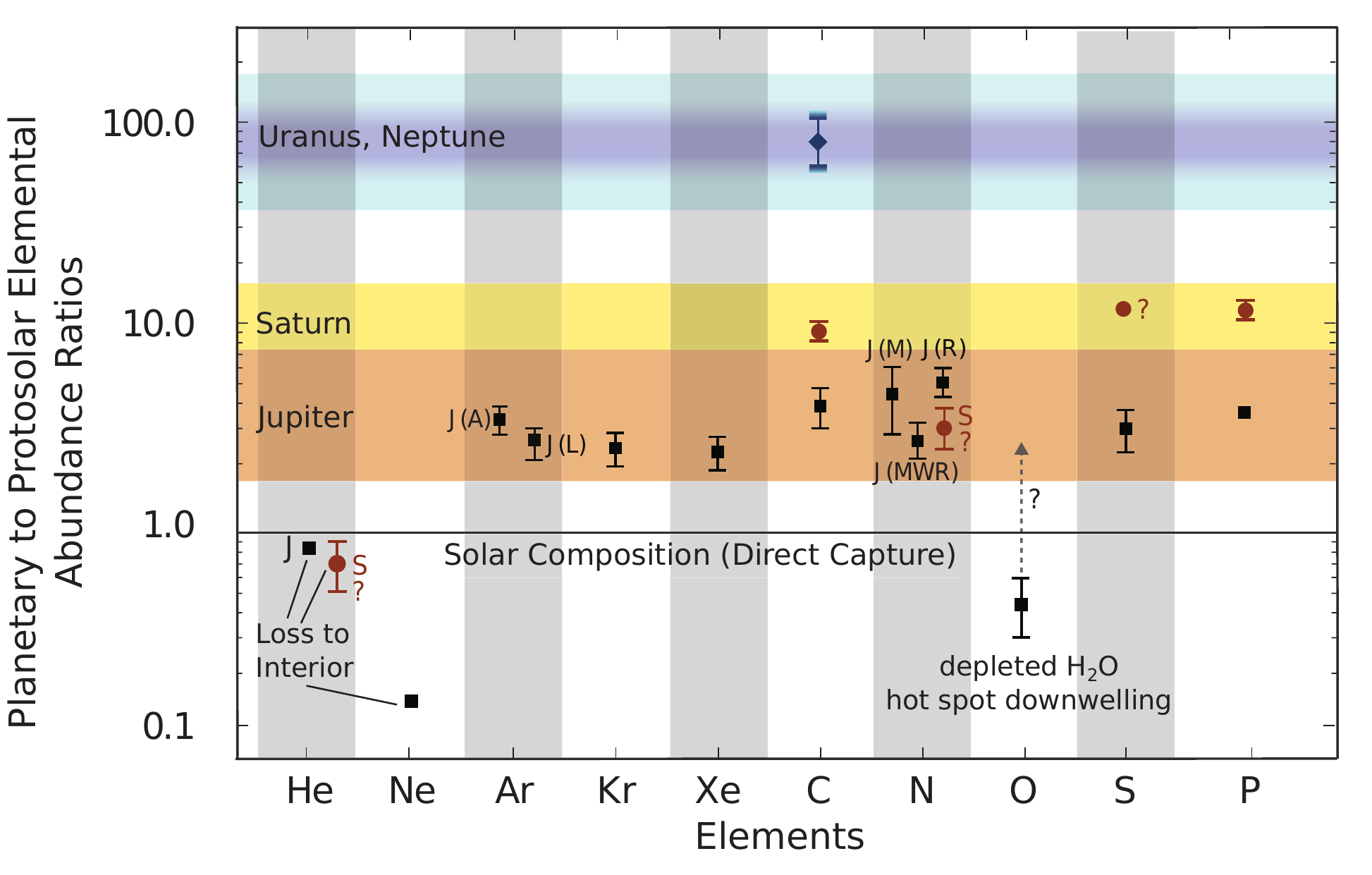}
\caption{Elemental abundance ratios in the atmospheres of Jupiter, Saturn, Uranus and Neptune. ``N'' in Jupiter represents values from ammonia (NH$_3$) abundance measurements made by the Galileo probe (mass spectrometer: [J(M)] and attenuation of probe radio signal: J[R]) and the Juno microwave spectrometer [J(MWR)], whereas ``Ar'' values are based on protosolar Ar/H values of \cite{As09} [J(A)] and \cite{Lo09} [J(L)]. Saturn’s He and N are labeled S. N/H of Saturn is a lower limit, and S/H is highly questionable. Only C/H is determined for Uranus and Neptune from ground-based CH$_4$, but remains uncertain. See \cite{At19} and references therein for all relevant details. [This version of the figure appears as Fig. 3 in \cite{At20}, this issue, which was adapted from Fig. 2.1 of \cite{At19}, with permission from Cambridge University Press, PLSclear Ref No: 18694].}
\label{plot0}   
\end{figure}

Figure \ref{plot0} shows the elemental abundance ratios in the giant planets relative to those in the proto-Sun, based on the values listed in Table \ref{val} (see \cite{At19} for details). For Uranus and Neptune, only the C/H has been directly determined from the analysis of Voyager radio occultation and Hubble Space Telescope data, respectively. The C/H ratio on these planets is inferred to be 80 $\pm$20 $\times$ solar, but is uncertain as it involves assumptions about the amount of H tied up in non-methane volatiles, whose abundances are unknown (see \cite{At20}, this issue, for details). NH$_3$ and H$_2$S would be expected to be significantly enriched relative to solar values, much like the observed enrichment in CH$_4$, assuming that the building blocks of Uranus and Neptune have solar C/N and C/S. Furthermore, S/N should also be solar, which would predict a value of $\sim$0.2.  

However, ground-based observations have measured ammonia in the atmospheres of Uranus and Neptune since the late 1970's. \cite{Gu78} found that unlike Jupiter and Saturn, NH$_3$ is subsolar by a factor of 100 at temperatures below 250 K (pressures $<$ 40 bars) in the observable atmospheres of these planets. This finding was later confirmed by Very Large Array (VLA) observations of radio emission spectra from Uranus and Neptune over the wavelength interval from $\sim$0.1 to 20 cm, corresponding to pressures down to $\sim$50 bars \citep{de89,de91}. At the wavelengths considered, the opacity from NH$_3$ is important enough, compared with those from H$_2$O and H$_2$S, to derive a subsolar NH$_3$ abundance in Uranus and Neptune. The observed depletion of NH$_3$ cannot be a consequence of condensation of this species, which is expected to occur at much shallower levels. The depletion of ammonia manifests itself in the presence of H$_2$S in the upper tropospheres of Uranus and Neptune. This is due to the fact that in thermochemical models of the giant planets NH$_3$ serves as a sink for H$_2$S by forming a cloud of ammonium hydrosulfide. With little ammonia available, H$_2$S vapor can survive. In fact, H$_2$S vapor has now been detected on both Uranus and Neptune  \citep{Ir18,Ir19a}. With S/N $\ge$5, an H$_2$S ice  would form at 3 bars or deeper. The presence of an opacity source in this pressure region was inferred from the VLA data \citep{de89,de91}. These observations appear to suggest non-solar C/N and S/N abundance ratios in the Ice Giants that may be tracers of building blocks that were also non-solar \citep{Ma19}. It should be stressed however that although NH$_3$ and H$_2$S have been detected in the tropospheres of Uranus and Neptune, their bulk abundances are unknown. It is possible that the NH$_3$ could be removed in a purported water ocean at the 10-kilobar level or deeper, and possibly an ionic ocean at 100 kilobars and deeper (see \cite{At20}, this issue, for details of cloud structure and ammonia depletion). In other words, the question of the possible N and S depletions in the atmospheres of Uranus and Neptune remains open.

\begin{table}
\caption[]{Elemental abundance ratios in the Sun, Uranus and Neptune (this is a subset of Table 1 in \cite{At20}}
\begin{center}
\begin{tabular}{lccc}
\hline
\hline
\noalign{\smallskip}
Elements		 	&  Sun-Protosolar$^{(a,b)}$ 	& Uranus/Protosolar 			& Neptune/Protosolar			\\	
\noalign{\smallskip}
\hline
\noalign{\smallskip}
He/H				& 9.55 $\times$ 10$^{-2}$		& 0.94 $\pm$ 0.16$^{(c)}$		& 1.26 $\pm$ 0.21$^{(c)}$			\\
				& 						& 						& 0.94 $\pm$ 0.16$^{(c)}$			\\
Ne/H				& 9.33 $\times$ 10$^{-5}$		& NA						& NA							\\
Ar/H				& 2.75 $\times$ 10$^{-6}$		& NA						& NA							\\
Kr/H				& 1.95 $\times$ 10$^{-9}$		& NA						& NA							\\
Xe/H				& 1.91 $\times$ 10$^{-10}$	& NA						& NA							\\
C/H				& 2.95 $\times$ 10$^{-4}$		& 80 $\pm$ 20$^{(d)}$		& 80 $\pm$ 20$^{(e)}$			\\
N/H				& 7.41 $\times$ 10$^{-5}$		& 0.01--0.001$^{(f)}$		& 0.01--0.001$^{(f)}$			\\
O/H				& 5.37 $\times$ 10$^{-4}$		& NA						& NA							\\
S/H				& 1.45 $\times$ 10$^{-5}$		& $>$ ($\sim$0.4--1.0)$^{(g)}$	&  $>$ ($\sim$0.1--0.4)$^{(h)}$		\\
P/H				& 2.82 $\times$ 10$^{-7}$		& NA						& NA							\\
\hline
\end{tabular}
\end{center}
($a$) Protosolar values based on the solar photospheric values of \cite{As09}. ($b$) Protosolar metal abundances relative to hydrogen can be obtained from the present day photospheric values \citep{As09}, increased by +0.04 dex, i.e. ~11\%, with an uncertainty of $\pm$ 0.01 dex; the effect of diffusion on He is very slightly larger: +0.05 dex ($\pm$0.01). ($c$) \cite{Ga95}; two values are given for Neptune, one without N$_2$ in the atmosphere (larger He/H) and the other including N$_2$ in order to explain presence of HCN. (d) \cite{Sr11}; E. Karkoschka and K. Baines, personal communication (2015). (e) \cite{Ka11}. (f) See text. (g) \cite{Ir18}, (h) \cite{Ir19a}; lower limit below an H$_2$S cloud, based on the detection of H$_2$S gas in the 1.2--3 bar region above the cloud. This S/H is not necessarily representative of the actual value in the deep well-mixed atmosphere (see text).
\label{val}
\end{table}

\section{Scenarios of Volatiles Delivery}
\label{scenar}

\begin{figure}
\center
\includegraphics[width=1\textwidth]{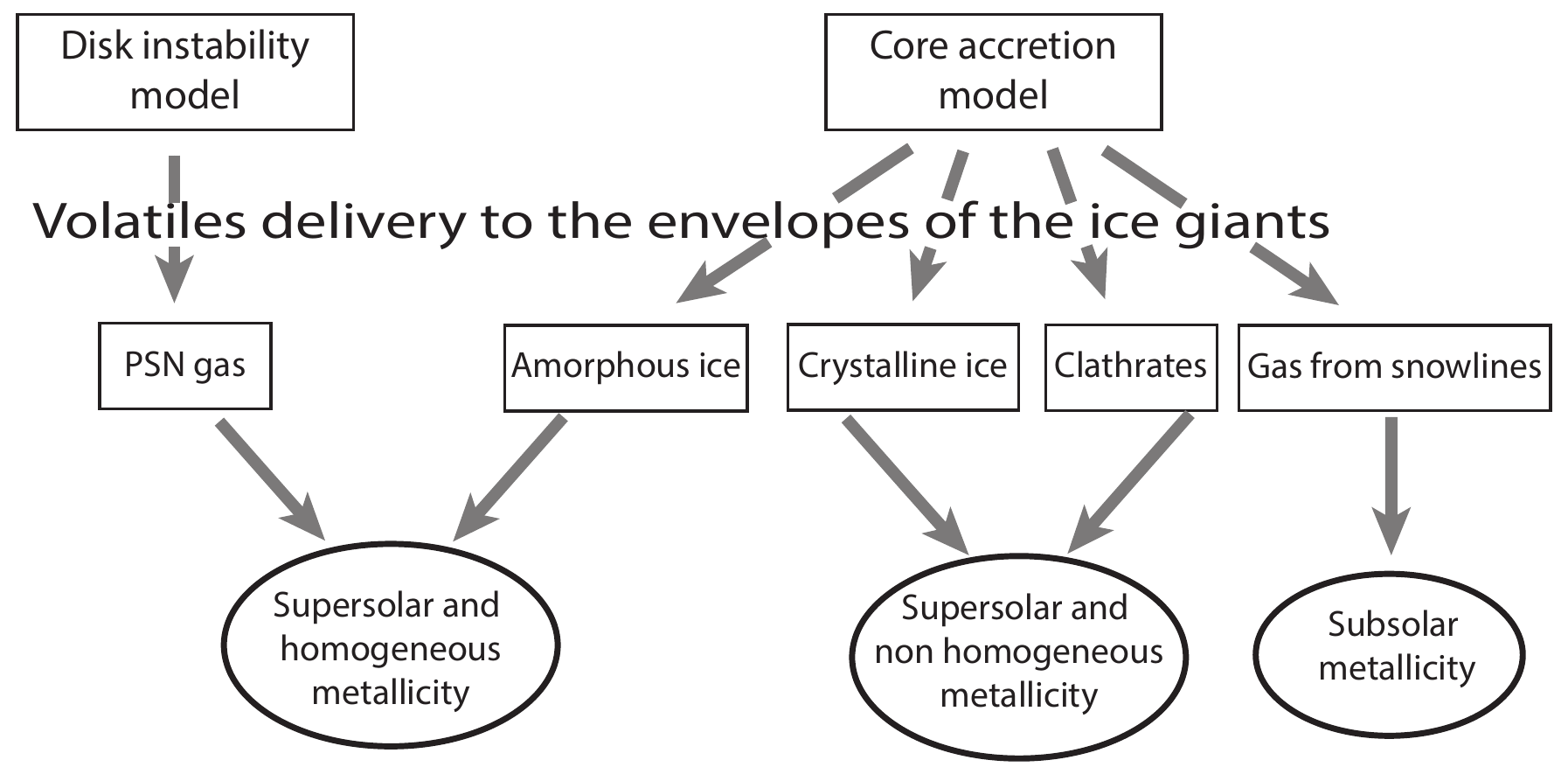}
\caption{Metallicity of the envelopes of Uranus and Neptune derived for each of the considered delivery mechanisms of volatiles.}
\label{tree}   
\end{figure}

In this section, we depict the two main mechanisms of giant planet formation, i.e. the disk instability and core accretion models. In the first scenario, giant planets essentially grow from gas while, in the second scenario, their accretion requires the formation of a solid core before the accretion of gas and alternately solids. For each of these scenarios, we review and investigate the different delivery mechanisms of volatiles that can account for the volatiles enrichments observed in the ice giants atmospheres. Figure \ref{tree} summarizes the implications for the metallicity of the envelopes of Uranus and Neptune in the cases of the delivery mechanisms of volatiles discussed below.

\subsection{Disk Instability Model}

One model for the formation of the giant planets posits the destabilization of the disk at an early stage in its evolution. Spiral density waves develop which transport angular momentum outward and mass inward, the latter allowing the self-gravity of the disk to tap into the free energy of the combined central protosun-disk system. A disk will become gravitationally unstable as the amplitude of the spiral density waves increases, if the disk is sufficiently massive and sufficiently cool \citep{Ar10}. It will then break up into discrete mass concentrations which can continue to compact under their own self gravity, leading to a large number of giant planets \citep{Bo97}. The instability mechanism is fast, leading to giant planet formation in hundreds to perhaps thousands of years \citep{Ma04}. Subsequent interaction with the disk can lead potentially to large enrichments of heavy elements, even large cores \citep{Bo11}. Subsequent erosion of the H-He envelope might then produce something looking like an ice giant. Therefore, it may be difficult to use heavy-element composition to distinguish between core accretion and disk instability models since both cases predict significant volatile enrichments in the envelope.

However, it may be difficult for most disks to undergo the fragmentation required to generate giant planets even if the disk instability mechanism is at play. Extraction of gravitational potential energy as spiral density waves moves angular momentum outward and mass inward leading to heating of the disk, which tends to bring the disk away from the criterion of instability defined in \cite{Ar10}. The production of multiple giant planets in orbital relationships that are dynamically unstable leads to ejection of some planets and resulting placement of others on eccentric orbits that seem inconsistent with the architecture of the outer solar system. Mechanisms for post-fragmentation atmospheric enrichment that lead, for example, to the variation in enrichment seen in Jupiter have yet to be demonstrated. Finally, whether enough solid material remains to subsequently build the terrestrial planets, and even to populate the outer solar system with the initial reservoir of solid bodies that must be the source of the Oort Cloud and primordial Kuiper Belt, are open questions. The disk instability model cannot be ruled out as the operative mechanism for the formation of the giant planets, including Uranus and Neptune, but seems to raise more questions than it answers. 

\subsection{Core Accretion Model}

In the core accretion scenario, the formation of giant planets starts with the build-up of a core, followed by the slow contraction of a gaseous envelope, and, if the envelope becomes massive enough, a phase of rapid, so-called runaway, gas accretion is eventually triggered \citep{Po96}. Giant planets formation is then first dominated by the accretion of solids and later on, by the accretion of gas. In this respect, the bulk composition of Uranus and Neptune, with inferred metallicities in the range $\sim$0.8--0.9 \citep{He11,Po19}, is indicative of the fact that these planets never reached the runaway gas accretion phase, or did so only towards the very end of the lifetime of the PSN, thus preventing them from becoming gas dominated planets like Jupiter and Saturn. 

In details, the composition of the envelopes of Uranus and Neptune depends on their formation history, the size and composition of their building blocks, and the metallicity of the gas they accreted. The solids accreted by Uranus and Neptune could have either sunk to the center of the planets or been diluted in their envelope depending on their sizes \citep{Po88,Lo17}. The formation timescale of Uranus and Neptune from large ($>$ km sized) planetesimals would by far exceed the expected lifetime of the PSN \citep{Le07,Le10}. Instead, these planets most likely grew by accreting millimetre to centimetre sized drifting particles in a process known as pebble accretion \citep{La12,La14}. The pebbles are more likely to be diluted in the envelopes of the planets rather than to sink down to a core \citep{La14,Ch17,Br18}. Also, contrary to large planetesimals whose composition is somewhat locked up once they have formed, the volatile composition of the pebbles is expected to quickly adapt to the local temperature and pressure conditions of the disk, with condensation and evaporation taking place around the location of the so-called snowlines \citep{Ro13,Al14,Bo17}. Pebbles are also able to transport volatile species and noble gases that could have been trapped within the amorphous matrix of water ice \citep{Mo19} or eventually in the form of clathrates. Finally, it should be noted that the high obliquities of Uranus and Neptune might be the result of violent impacts that took place during the final assembly of the ice giants \citep{Mo12,Iz15,Re19}. Such giant impacts could not only deliver some heavy elements to the envelope of the planets, but also substantially alter their interior and thus affect their long-term evolution \citep{Re19,Li19}.

In this context, a long-standing debate is the question of the nature of the reservoirs of materials that contributed to the formation of giant planets, comets, and Kuiper-Belt Objects \citep{Ow99,Ga01,Ba07,Ru15,Mo18}. In the following, we describe the different delivery mechanisms of the volatiles to the forming giant planets in the framework of the core accretion model.

\subsubsection{Amorphous Ice}
\label{amorphous}

\begin{figure}
\center
\includegraphics[width=0.9\textwidth]{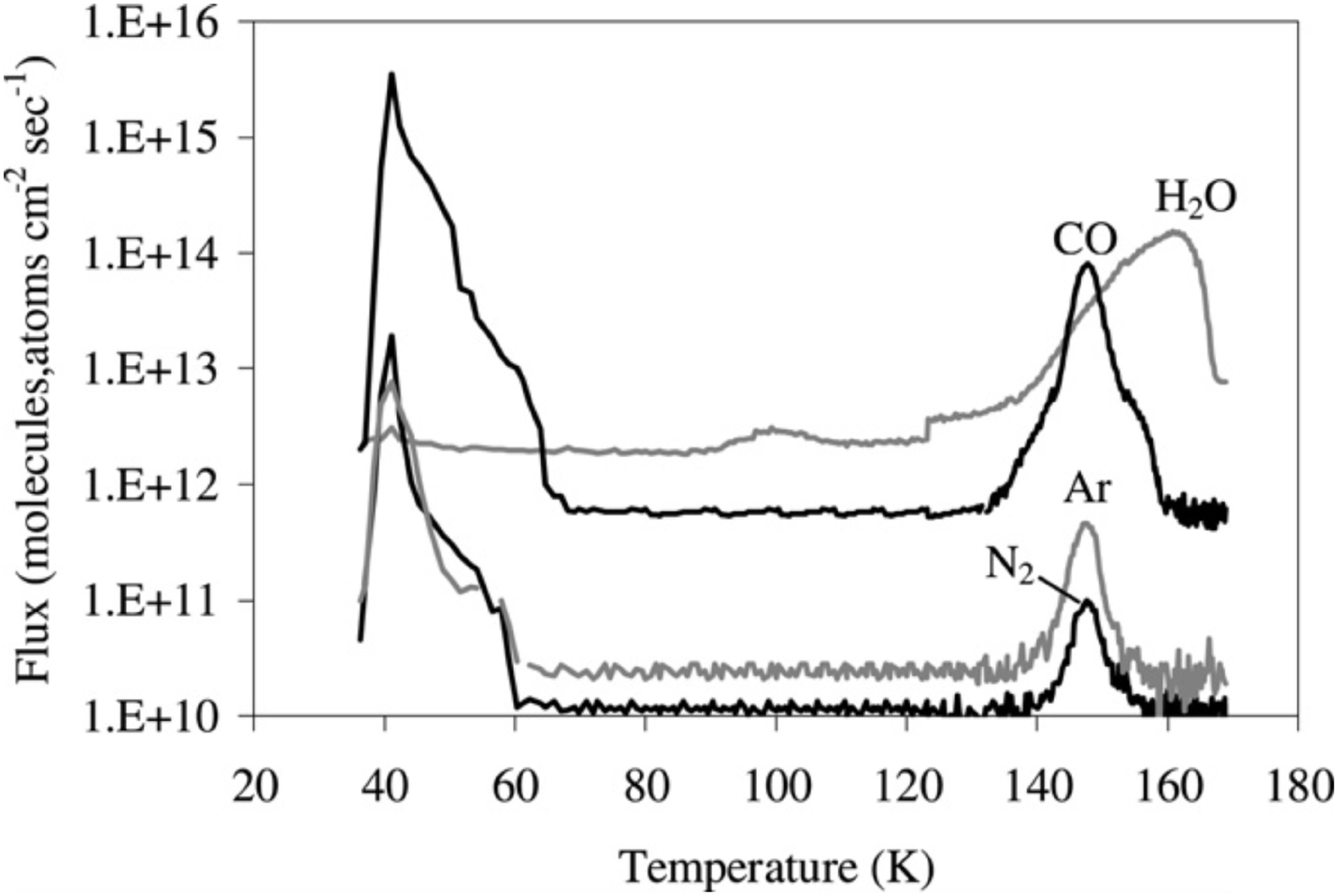}
\caption{Flux of gases vs temperature in the experiments of Bar-Nun and colleagues. Low temperature (40--60 K) release is of pure ices sublimating off of grains. Release between 140--160 K is the conversion of amorphous to crystalline ice. [Figure 1 of \cite{Ba07}, with permission from Elsevier, XXX].}
\label{plot_bar}   
\end{figure}

The observed uniform enrichment in Jupiter in a class of heavy elements not affected by deep envelope miscibility (He, Ne) is puzzling, given that the original molecular carriers and the noble gases Ar, Kr, Xe have such widely varying volatilities. For example, in the case of carbon, the two major carriers CO and CH$_4$ vary in their vapor pressures at 40 K by three orders of magnitude.  One scenario invoked in the literature to account for this homogeneous enrichment is the delivery of amorphous planetesimals to Jupiter’s envelope \citep{Ow99}. The ratio of trapped volatiles to water in amorphous ice can be up to about 8\%, depending on its porosity and surface available for adsorption \citep{Sc89}. Figure \ref{plot_bar} shows the release of gas from amorphous ice versus temperature derived from experiments depicted in \cite{Ba07} for an initial deposition ratio of H$_2$O:CO:N$_2$:Ar = 100:100:14:1. Once the condensed ices of each species are sublimated below 60 K, a significant amount of CO and Ar are retained but very little N$_2$. The result, if Jupiter was indeed seeded directly by amorphous planetesimals, might be a deficit of nitrogen unless ammonia was also present in sufficient amounts in the planetesimals. Also, a rather large water enrichment --at least similar to that of carbon-- might result.
 
\cite{Ow99} proposed that the planetesimals that seeded Jupiter were extremely cold, perhaps 30 K, so that the entire trapped volatile load seen in Figure \ref{plot_bar} --directly condensed ices and that trapped in the amorphous water ice-- could have been delivered to Jupiter. This would produce a more uniform enrichment, as observed, but would require very cold conditions at Jupiter. This scenario is more appropriate in the cases of Uranus and Neptune, which were located farther out in the protoplanetary disk and plausibly accreted very cold planetesimals. Assuming the two planets i) never reached the pebble isolation mass, which prevents the efficient accretion of centimeter- to meter-sized solids \citep{Bi18}, and ii) the absence of larger planetesimals in the feeding zones, their growth from a mixture of gas and amorphous solids would lead to a uniform enrichment of heavy elements in their envelopes. If Uranus and Neptune exceeded the pebble isolation mass at the time their envelopes formed, one might expect a much more limited enrichment of heavy elements, but still uniform, unless large planetesimals were dominant in the feeding zones. In this case, a substantial and uniform enrichment of heavy elements would be expected in the envelopes since the accretion of such planetesimals would not be affected by the pressure bump defining the pebble isolation mass \citep{Bi18}. Either way, the signature of the accretion of amorphous solids would be the distinct difference in enrichment of water (hence, O) between the ice giants and Jupiter. 

\subsubsection{Crystalline Ices}
\label{cryst}

\begin{figure}
\center
\includegraphics[width=0.7\textwidth]{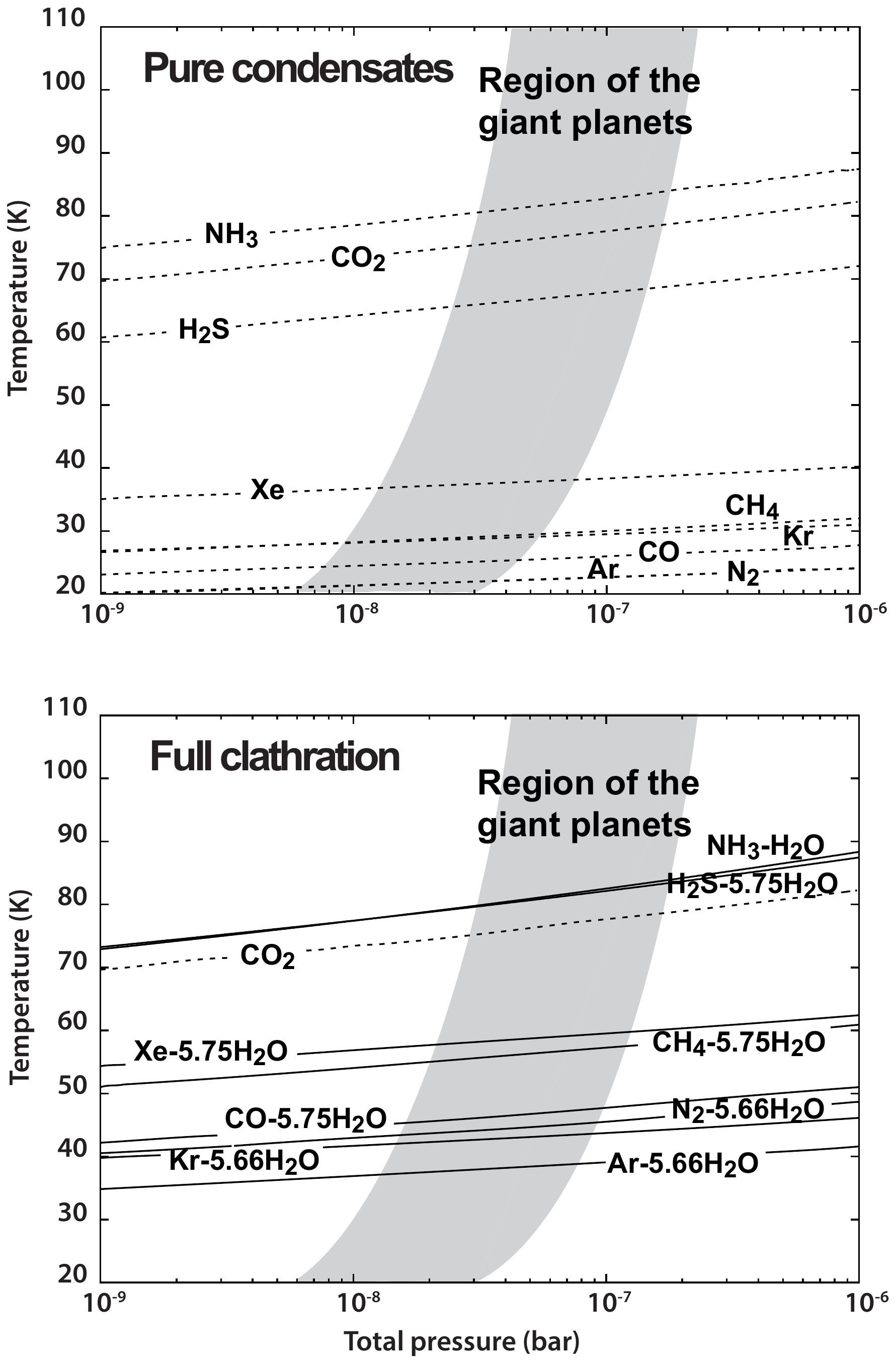}
\caption{Condensation sequence of ices and cooling curve of the PSN in the formation region of the giant planets. Top panel: equilibrium curves of pure condensates (dashed lines), assuming protosolar elemental abundances \citep{Lo09} and CO:CO$_2$:CH$_4$ = 10:10:1 and N$_2$:NH$_3$ = 10 in the gas phase of the disk. Species remain in the gas phase above the equilibrium curves. Bottom panel: same as top panel but in the case of hydrate (NH$_3$-H$_2$O) and clathrate formation (X-5.75H$_2$O or X-5.67H$_2$O; solid lines), and crystallization of pure CO$_2$ condensate (dotted line), assuming a full efficiency of clathration. CO$_2$ is the only species that crystallizes at a higher temperature than its associated clathrate in the pressure conditions of the PSN \citep{Mo09a}.}
\label{plot_clat}   
\end{figure}

When interstellar medium amorphous grains enter the PSN and cross the amorphous-to-crystalline transition zone (ACTZ), i.e. the zone within which the disk temperature reaches or exceeds $\sim$143 K \citep{Ba07}, water ice crystalizes and the adsorbed volatiles are released to the disk gas phase. Because the PSN slowly cools down with time, the released volatiles will form crystalline ices whose condensation temperatures are lower than the one needed to crystallize amorphous ice. These crystalline ices can consist in pure condensates forming in a temperature range comprised between $\sim$20 and 150 K in the PSN, depending on their abundances relative to H$_2$ and the equilibrium pressures of the considered species (see Fig. \ref{plot_clat}, top panel). Alternatively, crystalline ices can exist in the form of clathrates, which are ice-like inclusion compounds forming with nonpolar guest molecules surrounded with hydrogen-bonded water cages \citep{Sl08}. The ratio of trapped volatiles to water in clathrates is up to about 1:6 total trapped gas \citep{Sl08}. Clathrates typically crystallize in the PSN at higher temperatures than those of the pure condensate forms of their encaged molecules (Fig. \ref{plot_clat}, bottom panel). Clathrate formation essentially depends on the availability of crystalline water ice in the PSN \citep{Lu85,Mo10}, and also on the kinetics of entrapping, which remains to be assessed \citep{Gh19a,Gh19b,Ch19}. The building blocks of all giant planets of our solar system, as well as comets and Kuiper-belt objects, might have formed from one of these two kinds of crystalline ices, or maybe from a mixture of both. 

Assuming formation at temperatures low enough to enable the crystallization of ultra-volatiles such as Ar or N$_2$ (see Fig. \ref{plot_clat}, top panel), planetesimals agglomerated from pure condensates should display a composition slightly different from the one derived from a protosolar mixture. Because the equilibrium temperatures of the considered ices vary over several dozens of kelvins, they are expected to condense at different epochs of the PSN evolution. As a result, the volatile A/volatile B abundance ratio in solid phase should be higher than the one derived from their gas phase abundances if volatile A crystallizes at a higher temperature than volatile B. This is due to the fact that the disk dissipates with time, implying a continuous decrease of its surface density (see Table 2 of \cite{Mo09a} as an illustration of this effect). On the other hand, the composition of planetesimals agglomerated from clathrates depends on the availability of crystalline water at their crystallization epochs. If crystalline water is sufficiently abundant to trap all volatile species in presence, then the composition of the icy phase in planetesimals should also reflect a supersolar abundance of water. For the same reasons as those invoked in the case of pure condensates, the relative mole fractions among the encaged volatiles should slightly depart from protosolar, because of the continuous decrease of the disk’s surface density at the different epochs of entrapping. 

\begin{figure}
\center
\includegraphics[width=0.6\textwidth]{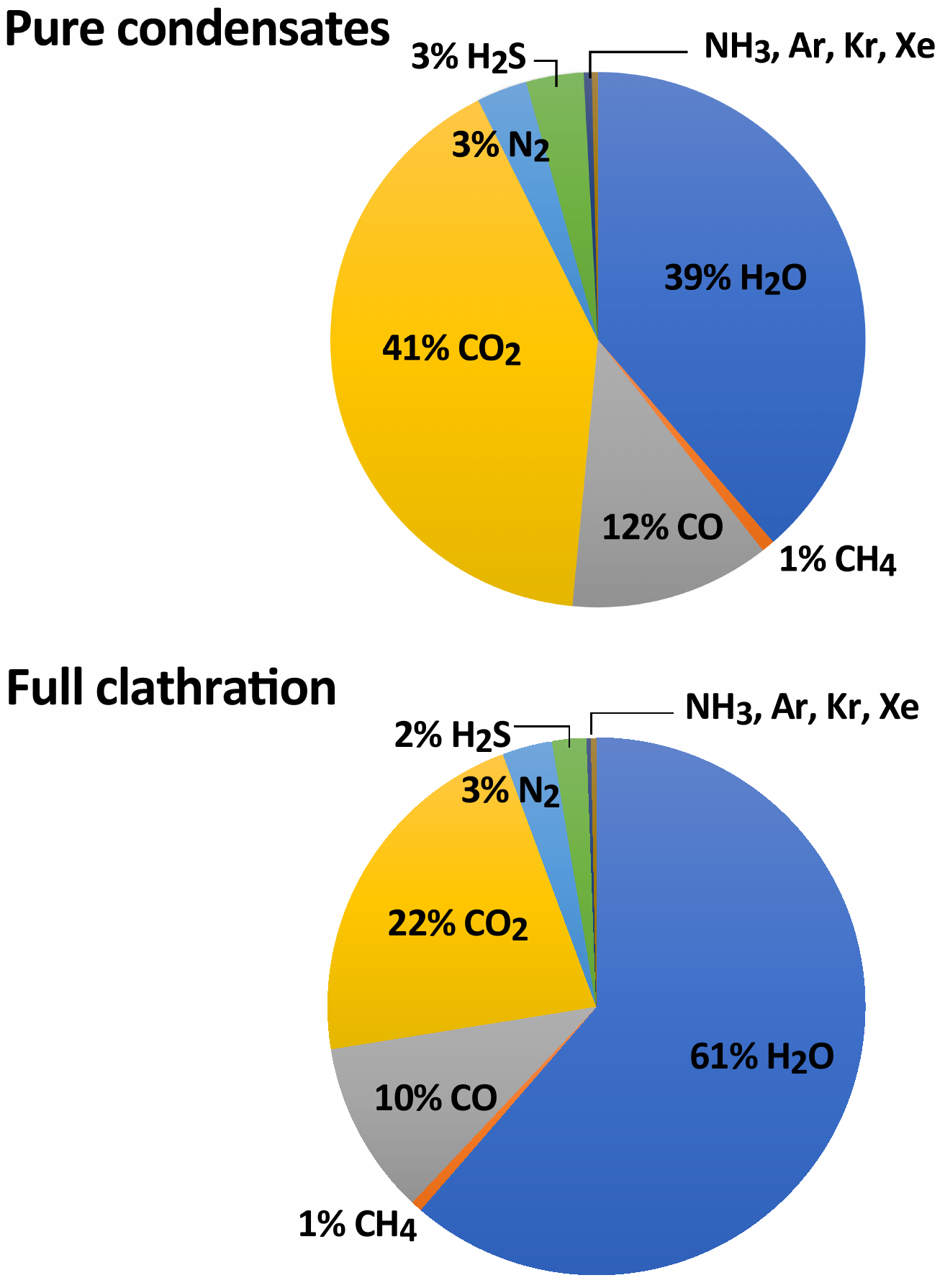}
\caption{Composition of the icy phase incorporated in solids resulting from the agglomeration of pure condensates (top panel) and clathrates (bottom panel). Gas phase and condensation/trapping conditions are those detailed in the caption of Fig. \ref{plot_clat}.}
\label{plot_pie}   
\end{figure}

If i) the budget of crystalline water is not significant enough to enable the enclathration of all the volatiles present in the gas phase or ii) the disk temperature never reaches those needed for trapping the most volatile species, then the composition of clathrate must be calculated assuming the simultaneous entrapment of all guests present in the coexisting gas phase. The composition of such multiple guest clathrates is predicted via a statistical thermodynamics approach, based on the van der Waals-Platteeuw statistical theory and the derivation of interaction potential parameters from experimental data \citep{va59,Lu85,Mo10}. These models predict an efficient trapping of CO, H$_2$S, Kr and Xe in clathrates at the expense of N$_2$ and Ar, assuming a PSN protosolar gas. They have been successfully used to interpret the composition of comet 67P/Churyumov-Gerasimenko (hereafter 67P/C-G), which has been found to be substantially depleted in Ar and N$_2$ compared to the protosolar values by the Rosetta/ ROSINA measurements \citep{Mo16,Mo18}. One direct consequence of the two aforementioned mechanisms, i.e. condensation and clathration, is that regions where the PSN temperature never reaches extremely low values cannot be populated by ultravolatile-rich planetesimals. The apparent deficiency of Saturn’s moon Titan in primordial CO, N$_2$ and Ar \citep{Ni05} could be thus interpreted as the consequence of its agglomeration of building blocks presenting such deficiencies in ultravolatiles \citep{He04,Al07,Mo09b}. Figure \ref{plot_pie} shows pie charts summarizing the composition of the icy phase incorporated in planetesimals assuming i) the crystallization of pure condensates from a PSN protosolar gas, and ii) the full clathration of volatiles. The case of full clathration of volatiles requires an oxygen abundance that is $\sim$1.7 times higher than the protosolar value ((O/H$_2$)$_\odot$ = 1.21 $\times$ 10$^{-3}$; \cite{Lo09}) for the adopted PSN gas phase mixing ratios (CO:CO$_2$:CH$_4$ = 10:10:1 and N$_2$:NH$_3$ = 10).

\subsubsection{Desorption of Volatiles at the ACTZ Snowline}
\label{ACTZ}

\begin{figure}
\center
\includegraphics[width=0.9\textwidth]{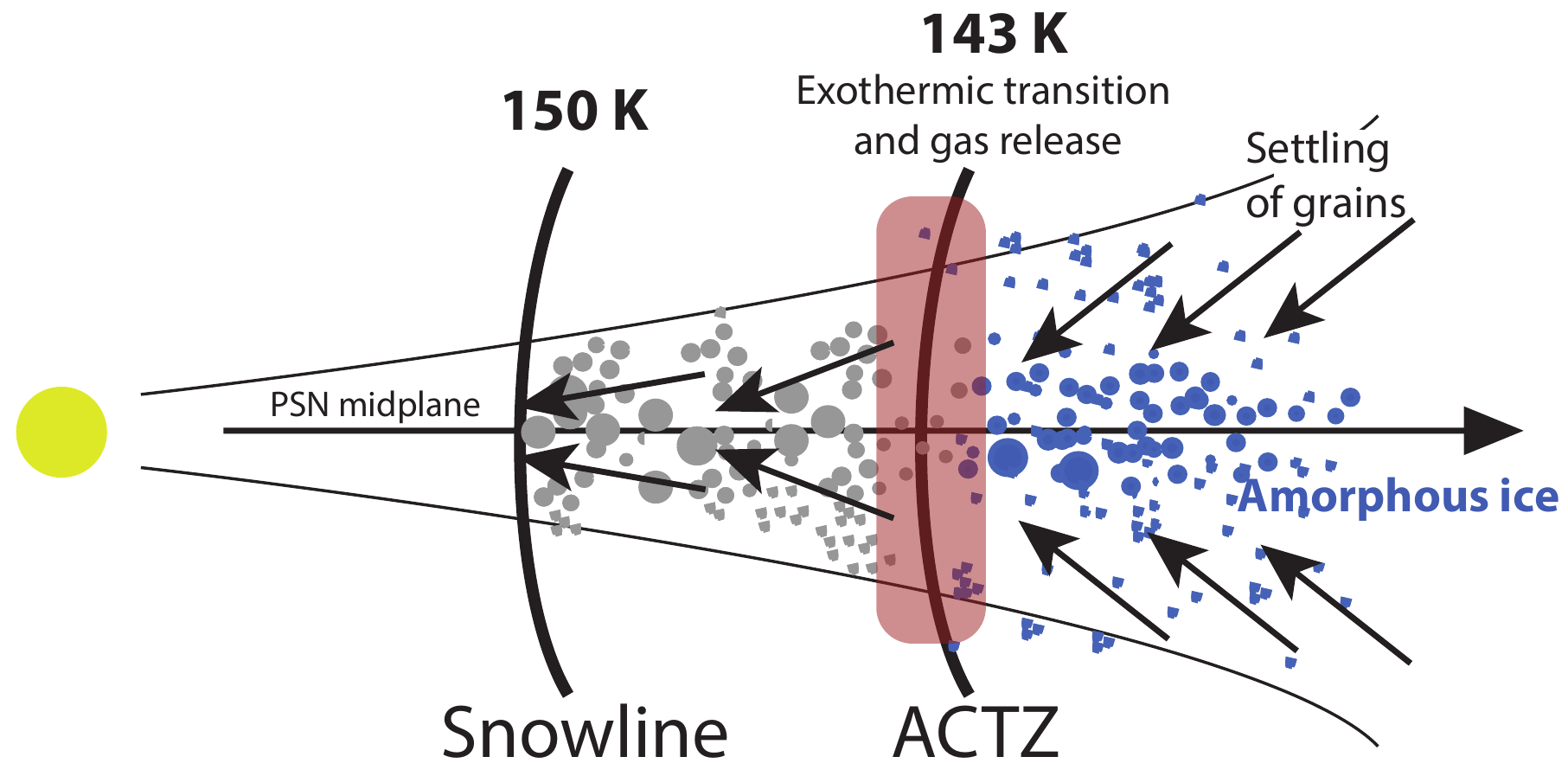}
\caption{Schematic of the relative positions of protoplanetary disk regions where amorphous ice is stable (blue), crystalline ice is present (grey) and then water-vapor only. The dividing lines between the amorphous and crystalline region is the ACTZ, that between ice and vapor is the snowline. Both are temperature-driven. Ice will occur inward of the snowline because inward migration is faster than sublimation for some particle sizes.}
\label{plot4}   
\end{figure}

In the core accretion model, giant planets grow over a timescale comparable to or longer than nebular thermal evolution. However, planetesimals may migrate radially on much shorter timescales, such that at some point during the growth of core and envelope, planetesimals formed at large distances and thus made of amorphous ice will cross the ACTZ while giant planets are growing (Fig. \ref{plot4}). In this scenario, when particles cross the ACTZ, volatiles desorb from amorphous ice and contribute to the disk’s gas phase \citep{Mo15}. The continuous release of vapors by the drifting particles at the location of the ACTZ creates a local enhancement of the gaseous abundances of the released volatiles, compared to their initial values \citep{Mo19}. The gas in the region between the ACTZ and snowline is charged with a supersolar abundance of species more volatile than water ice, but not with water vapor. This may provide an explanation for why the gaseous envelope of Jupiter is contaminated with several times solar abundances of heavy noble gases, carbon and nitrogen-bearing species, while the abundance of water from Juno data seems limited to less than twice solar \citep{Wa16}. So long as Jupiter formed between the ACTZ and the snowline, it could accrete a gaseous phase enriched in heavy elements except water ice. To limit the accretion of water ice one must then invoke the so-called pebble isolation mass, that of the growing giant planet at which the surrounding gaseous disk is sufficiently perturbed to limit the introduction of small planetesimals (pebbles) in to the giant planet \citep{Bi18}. 

\begin{figure}
\center
\includegraphics[width=0.9\textwidth]{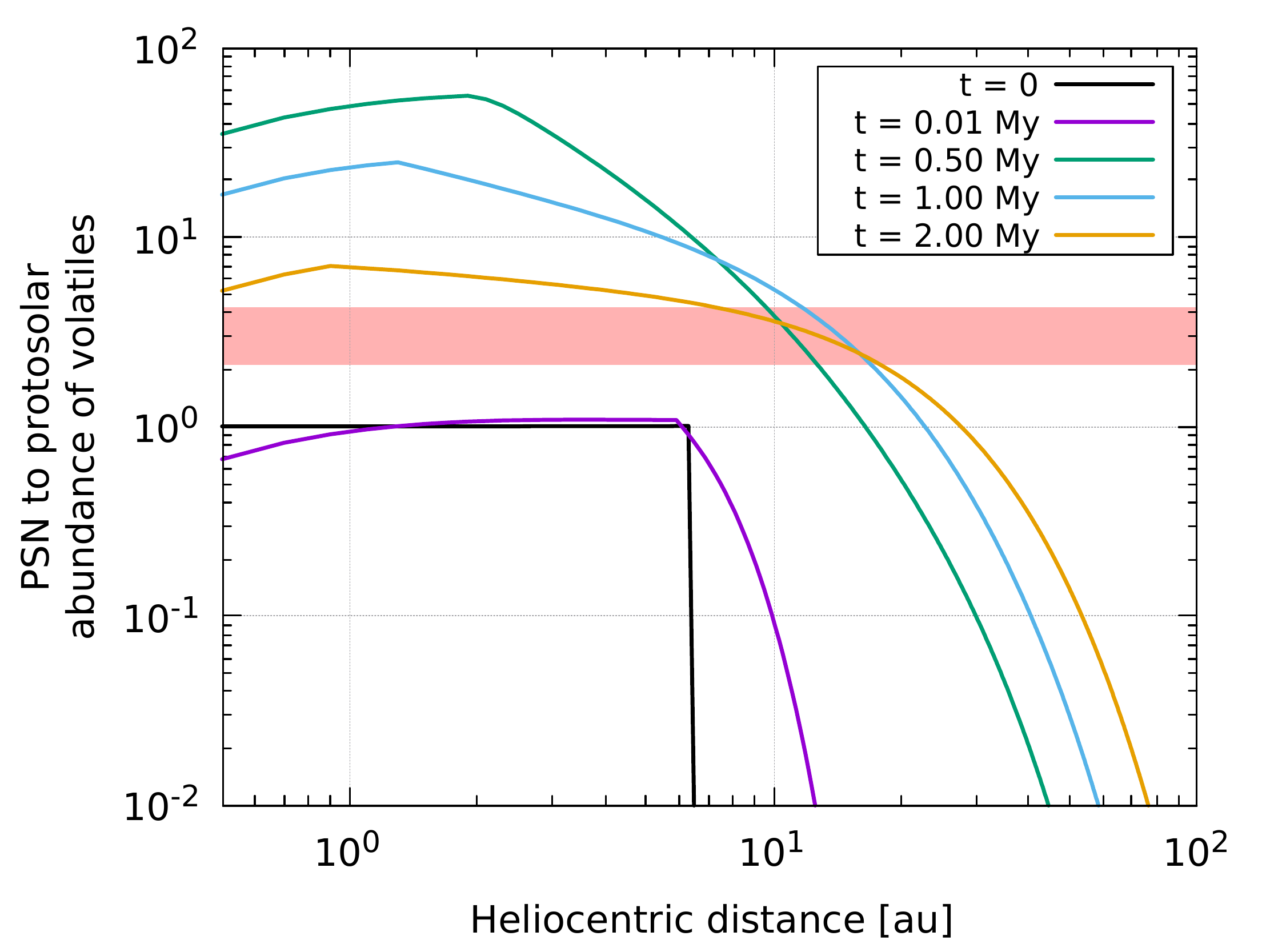}
\caption{Time and radial evolution of the abundances of volatiles released to the PSN gas phase by the icy grains subsequent to their drift through the ACTZ. Calculations have been performed for $\alpha$ = 5 $\times$ 10$^{-3}$ with the model detailed in \cite{Mo19}. The horizontal bar represents the range of volatile enrichments (nominal values) measured in Jupiter. With time, the metallicity of the PSN matches Jupiter's value at decreasing heliocentric distances and becomes progressively subsolar at radii located beyond.}
\label{plot5}   
\end{figure}

If the interplay between gas and solids as outlined above is responsible for the pattern of heavy elements observed in Jupiter, the contribution of this mechanism to the delivery of volatiles to Uranus and Neptune should be very moderate. Figure \ref{plot5} represents the evolution of the PSN metallicity at different epochs of its evolution. As the disk cools with time, the ACTZ moves inward, and volatiles that are continuously released at the ACTZ diffuse to the outermost regions of the PSN. The figure shows that i) Jupiter's metallicity is matched over a wide range of distances during the disk evolution, and ii) the disk metallicity drops down to subsolar values at larger heliocentric distances. The abundance of gaseous water is also expected to be by far subsolar since it is located well beyond its corresponding snowline in the formation region of Uranus and Neptune. Figure \ref{plot6} also illustrates this effect by showing the time evolution of C and O radial profiles in the PSN, assuming that CO and H$_2$O are the main carriers of these two elements, and in the case of the inward drift of pure condensates (see Section \ref{PC}). Given the fact that these two planets likely formed at the very end of the disk evolution, the abundances of the various gases dropped to even lower values than those reached at earlier epochs in their formation region.

\subsubsection{Enhancements of Vapors in the Vicinity of the Pure Condensates Snowlines}
\label{PC}

PSN gas can also be enriched in heavy elements via the vaporization of pebbles or particles when they cross the snowlines of pure condensates in the 20--40 K range \citep{Bo17}. The main differences with the ACTZ snowline is the presence of multiple snowlines whose number in the PSN corresponds to that of the considered volatiles and the adding of a sink term related to the condensation of vapors when they diffuse back to their respective snowlines. For the sake of illustration, we present some preliminary calculations based on a disk model whose structure and evolution is ruled by the following second-order differential equation \citep{Li74}:

\begin{eqnarray}
\frac{\partial \Sigma_g}{\partial t} = \frac{3}{r} \frac{\partial}{\partial r} \left[ r^{1/2} \frac{\partial}{\partial r} \left( r^{1/2} \Sigma_g \nu \right)\right]. \label{eqofmotion}
\end{eqnarray}

\noindent This equation describes the motion of a viscous accretion disk of surface density $\Sigma_g$ with radial distance $r$ and dynamical viscosity $\nu$, assuming hydrostatic equilibrium in the $z$ direction, and considering the Sun has reached its final mass of $1.0$ M$_{\odot}$. We use the prescription of \cite{Sh73} for $\alpha$-turbulent disks to obtain the gas viscosity $\nu$, and the mid-plane temperature $T_d$ is computed following the approach of \cite{Hu05}. The initial condition is \citep{Li74}:

\begin{eqnarray}
\Sigma_g \nu = \frac{C}{3\pi} exp\left(-\left(\frac{r}{r_0}\right)^{2-p} \right), 
\label{selfsim}
\end{eqnarray}

\noindent where $C=M_{acc,0}$, is the mass accretion rate onto the central star at $r=0$, whose value is assumed to be $10^{-7.6}$ M$_{\odot}$ yr$^{-1}$ \citep{Ha98}. We choose $p=3/2$, a value corresponding to an early time disk. $r_0$ is adjusted to give a starting disk's mass of $0.1$ M$_{\odot}$.  The gas viscosity and midplane temperature are computed at each time step, and the disk is evolved with respect to Eq. \ref{eqofmotion}.

In this study, the H$_2$-He-dominated disk is uniformly filled with H$_2$O and CO, with O and C abundances assumed to be protosolar \citep{Lo09}. The condensation temperatures of these two species almost encompass the condensation temperatures of all main volatiles that may be present in the outer PSN (see Fig. \ref{plot_clat}, top panel). The surface density $\Sigma_i$ of a trace species $i$ (for its vapor and solid phase) is numerically evolved, solving the following 1D advection-diffusion equation:


\begin{eqnarray}
\frac{\partial \Sigma_i}{\partial t} + \frac{1}{r} \frac{\partial}{\partial r} \left[ r \left(\Sigma_i v_i - D_i \Sigma_g \frac{\partial}{\partial r}\left(\frac{\Sigma_i}{\Sigma_g}\right) \right)   \right] + \dot{Q}_i = 0, 
\label{advectiondiffusion}
\end{eqnarray}

\noindent where $v_i$ and $D_i$ are the radial velocities and diffusivities of species $i$, and $\dot{Q}_i$ is a source/sink term accounting for sublimation/condensation of vapor $i$ from solid grains. Vapor properties are those of the PSN gas. Grains' properties are computed using the two-population algorithm detailed in \cite{Bi12}, and we consider that dust grains' dynamical properties are those of the dominant species at each radius. Finally, the evaporation and condensation rates are computed following the approach of \cite{Dr17}.

\begin{figure}[!h]
\centering
\includegraphics[width=0.8\textwidth]{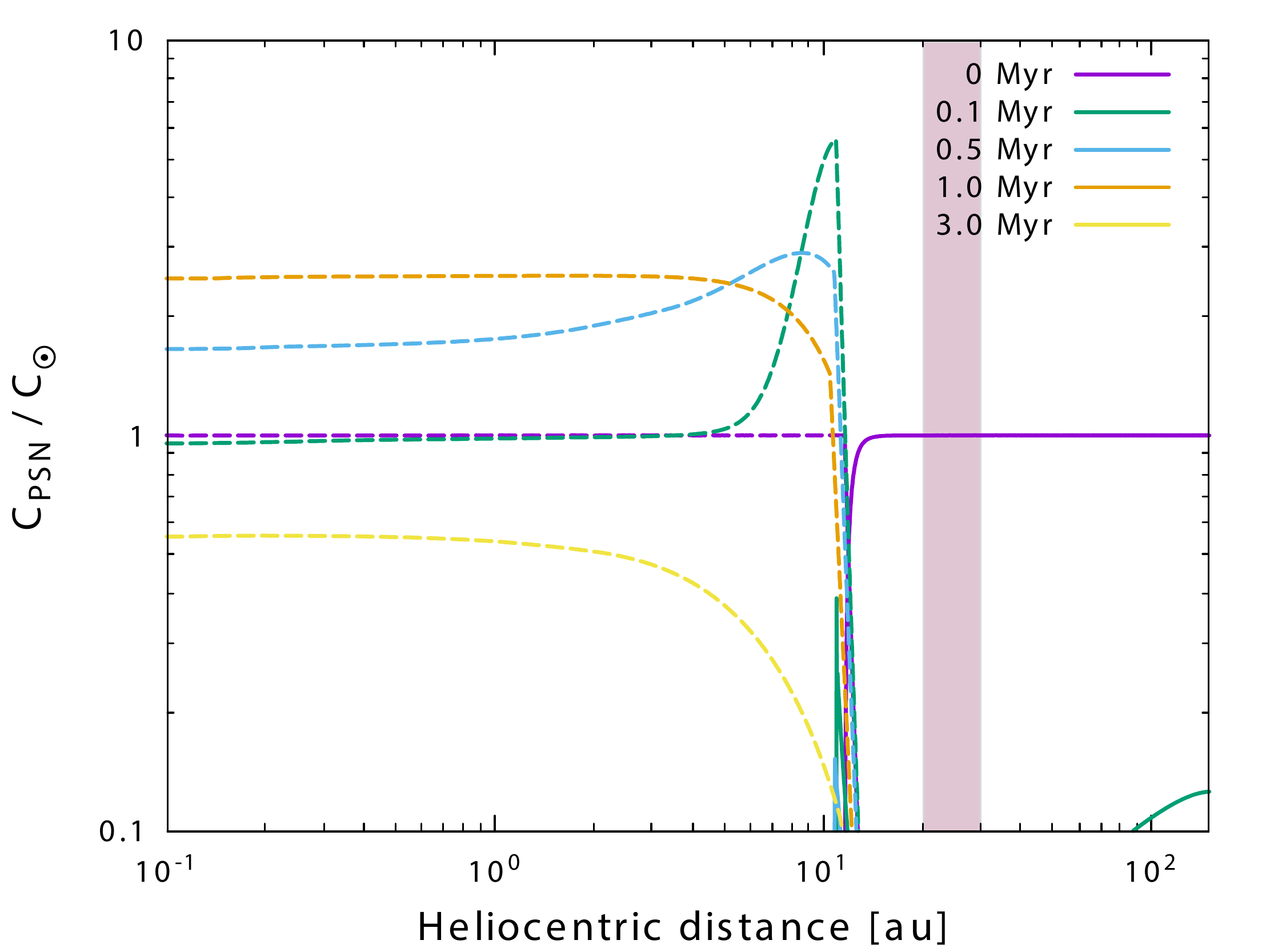} \\
\includegraphics[width=0.8\textwidth]{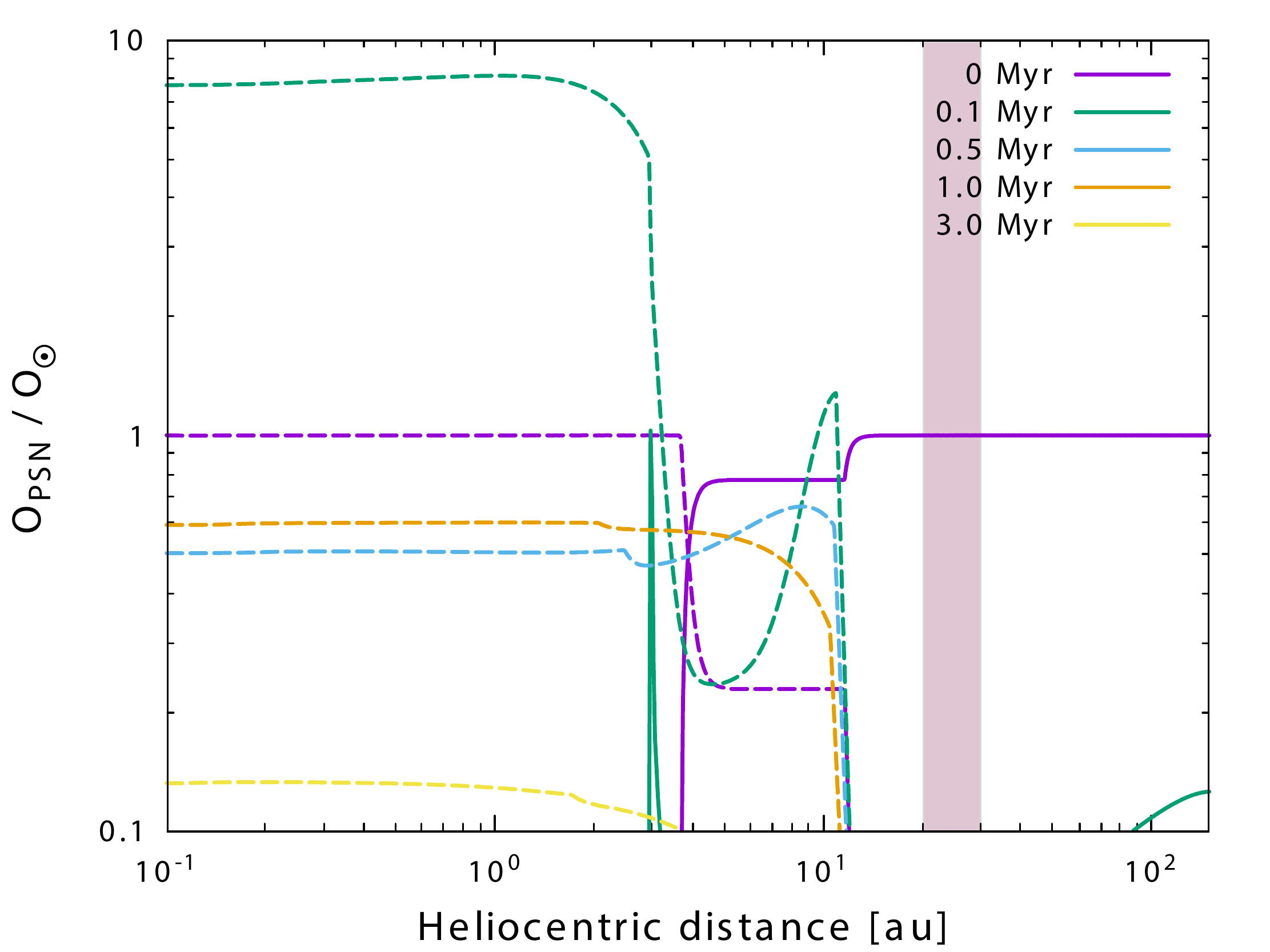}
\caption{Time evolution of radial profiles of C (top panel) and O (bottom panel) in the PSN, normalized to their respective protosolar abundances \citep{Lo09}. These two elements are assumed to be distributed between CO and H$_2$O who form pure condensates in the outermost regions of the PSN. Solid and dashed lines are used to identify the trace species in solid and gaseous phases, respectively. The vertical purple bars indicate the region of the ice giants in the solar system.}
\label{plot6}
\end{figure}

Figure \ref{plot6} shows that each pure condensate in solid grain form crossing its corresponding snowline sublimates and generates supersolar enrichments for the vapor released at this location. Because the vapor of a given species diffusing outwards recondenses back into solid grains, this prevents its presence far beyond its snowline. This effect contrasts with the scenario of volatiles released into the PSN when grains cross the ACTZ, and whose vapors cannot be trapped back when volatiles diffuse outward (Fig. \ref{plot5}). After 1 Myr of disk evolution, the gas phase abundances of C-- and O--bearing volatiles become substantially subsolar in the outer PSN, indicating that this effect cannot explain the supersolar metallicities measured in Uranus and Neptune. Because Uranus and Neptune likely formed later than this epoch, we find that i) the contrast between our results and the observations becomes reinforced, and ii) this mechanism is even less efficient than the one depicting the delivery of volatiles through the ACTZ.

One should mention that the scenarios of volatiles delivery through the ACTZ \citep{Mo19} or through the different snowlines of pure condensates, as shown in this paper, are based on dynamic models depicting the time evolution of the surface density, radial velocity and temperature radial profiles in the PSN, in contrast with  \cite{Al14} and \cite{Ob19} who use static disks. Computing the evolution of the PSN is an important step, since evolution is fast at early times, impacting both dust transport, determined by the PSN density and gas velocity, and thermodynamic conditions, fixed by midplane pressure and temperature.

\section{Atmospheric Signatures}
\label{Feat}

\begin{figure}
\center
\includegraphics[width=0.8\textwidth]{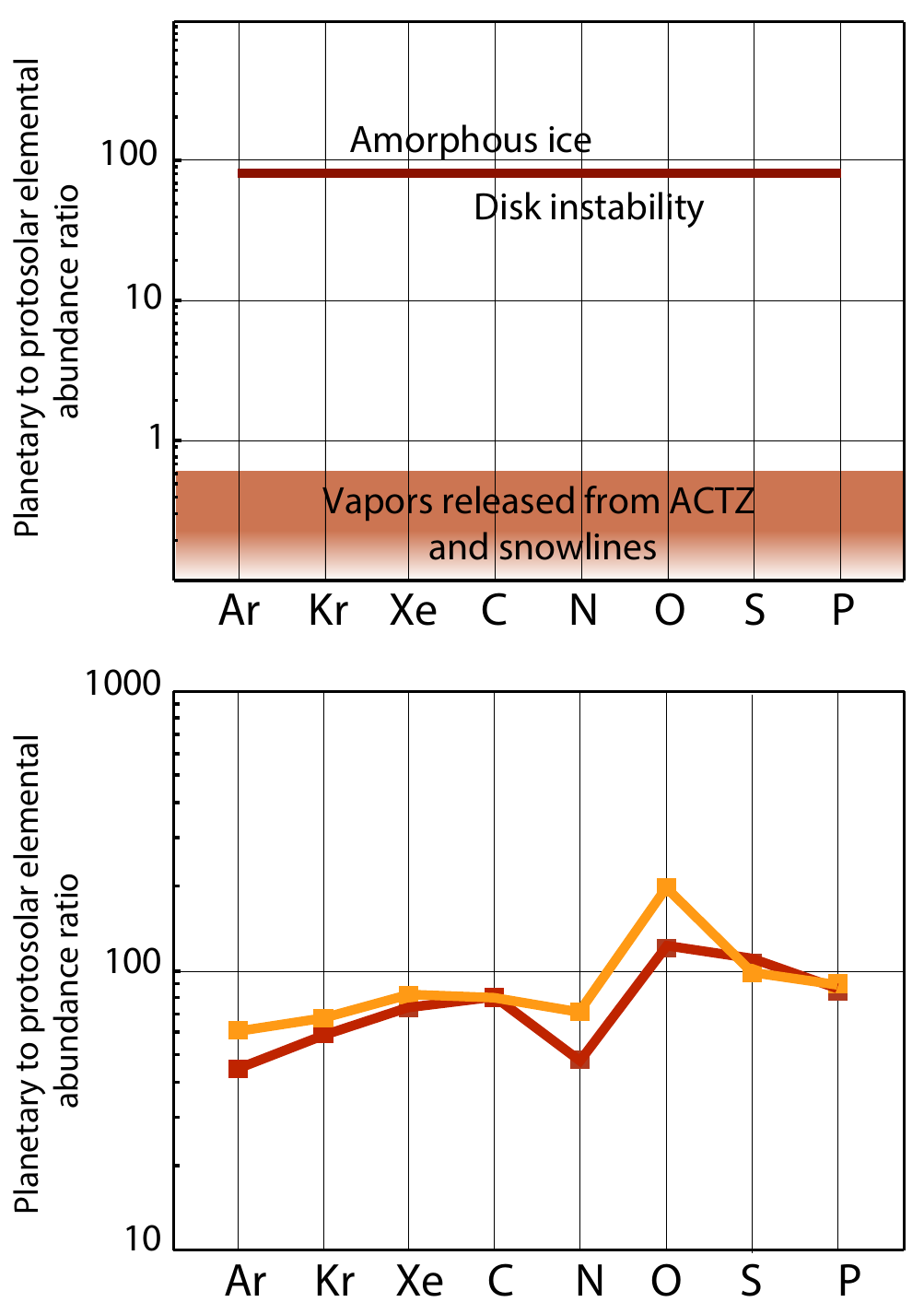}
\caption{Signatures of the different scenarios of volatiles delivery in the envelopes of Uranus and Neptune, assuming homogeneous mixing and that the measured C abundances are representatives of the bulk values. Calibration is done on a C abundance assumed to be 80 times protosolar (see Table \ref{val}). {\it Top panel}: volatiles delivered via disk instability or amorphous planetesimals in the framework of the core accretion model display significant supersolar and homogeneous volatiles enrichments, compared to their protosolar abundances. Volatiles delivered as vapors desorbed from the ACTZ or resulting from the sublimation of pure condensates at their respective snowlines display subsolar abundances in the envelopes. {\it Bottom panel:} atmospheric signatures of volatiles accreted in the ice giants in forms of pure condensates (red lines) or clathrates (orange lines).}
\label{plot7}   
\end{figure}

Here we use the characteristics of the discussed scenarios of volatiles delivery to derive their corresponding fingerprints in the atmospheres of Uranus and Neptune. To do so, all predictions of the volatile enrichments have been calibrated on the nominal C abundance measured in both planets ($\sim$80 times protosolar) (see Sec. \ref{abund}). Top panel of Fig. \ref{plot7} represents a homogeneous supersolar enrichment pattern in Uranus and Neptune adjusted to their C abundance measurements. This pattern is valid in the cases of i) volatiles delivered via disk instability, or ii) trapped in amorphous planetesimals, assuming the planets formed via core accretion. Top panel also shows the predictions of subsolar abundances of volatiles delivered by the release of vapors at the location of the ACTZ or at the position of the different snowlines of pure condensates. None of these two scenarios is able to predict the high C enrichment observed in Uranus and Neptune. The upper limits for the subsolar abundances are purely indicative. Much lower values can be obtained as a function of the utilized model, and also by assuming different formation epochs or distances. Bottom panel of Fig. \ref{plot7} represents the enrichment patterns of volatiles delivered in solids constituted of clathrates or pure condensates to the two ice giants. These enrichment patterns have been calculated with the model presented in \cite{Mo09b} who calculated the volatile enrichments in Jupiter and Saturn as a function of the efficiency of clathration in the PSN. The clathrate case predicts a water abundance in the ice giants that is $\sim$1.7 times higher than the value predicted from the pure condensates scenario (see Sec. \ref{cryst}). In the clathrate case, the Ar, Kr, Xe, C, N, O, S, and P abundances are predicted to be $\sim$60, 65, 79, 80, 68, 194, 95, and 87 times their protosolar abundances, respectively. In the pure condensates case, the Ar, Kr, Xe, C, N, O, S, and P abundances are predicted to be $\sim$44, 58, 72, 80, 47, 115, 103, and 81 times their protosolar abundances, respectively. Both cases are adjusted to the C measurement in the ice giants and assume it corresponds to the bulk value measured in the planets.  

Our predictions correspond to the assumption that the atmospheres of Uranus and Neptune are homogeneously mixed, and that no compositional gradient exists. This hypothesis is probably simplistic, given the fact that recent interior models suggest the presence of a compositional gradients in those planets \citep{He14,Ca17,Po19}. However, in absence of a proven mechanism depicting any variation among the relative abundances of heavy elements in the envelopes, the enrichment patterns acquired by the two planets should remain valid, even in the case of compositional gradient. In addition, if the atmospheric signatures result from the volatiles delivery in ices, these latter may have been i) either agglomerated by the ice giants during the growth of their envelopes or ii) accreted with the cores prior to their subsequent releases into the envelopes because of the erosion \citep{St82,Gu04,Wi12}. 

It is difficult to establish a link between planetary migration and the chemical composition of the ice giants. To the best of our knowledge, the only possible test would be the measurement of the noble gas isotopic ratios in the envelopes of ice giants. Since the ESA {\it Rosetta} spacecraft has measured a non-solar Xe mixture in comet 67P/C-G \citep{Ma17}, the identification of such a mixture in the atmospheres of the ice giants could tell if their building blocks came from the same formation location as the building blocks of 67P/C-G (see the discussion in \cite{Ma20}, this issue). These building blocks are supposed to have formed at a very high heliocentric distance in the PSN, may be the highest ever inferred, given the very high D/H ratio measured in the coma of 67P/C-G \citep{Al15}.

\section{Conclusions}
\label{Disc}

In this paper, we have investigated what could be the enrichment patterns of several delivery scenarios of the volatiles to the atmospheres of ice giants, having in mind that the only well constrained determination made remotely, i.e. the C abundance measurement, suggests that their envelopes possess highly supersolar metallicities, i.e. close to two orders of magnitude above that of the PSN. In the framework of the core accretion model, only the delivery of volatiles in solid forms (amorphous ice, clathrates, pure condensates) to these planets can account for the apparent supersolar metallicity of their envelopes. In contrast, all mechanisms invoking the delivery of volatiles in vapor forms, because of the inward drift of icy particles through various snowlines, predict subsolar abundances in the envelopes of Uranus and Neptune. Alternatively, even if the disk instability mechanism poses many questions in terms of feasibility in our solar system, it may be consistent with the supersolar metallicities observed in Uranus and Neptune, assuming the two planets suffered subsequent erosion of their H-He envelopes.

Atmospheric entry probes equipped with high-resolution mass spectrometers are the ideal tool to measure the abundances of the main species and their isotopes in the ice giants, given the difficulty of  remotely sounding the atmospheres down to the different condensation levels. We refer the reader to the works of \cite{At20} and \cite{Atk20}, this issue, for thorough discussions about the depths that can be sampled by a probe or an orbiter as a function of the science objectives.
Interestingly, subsequent probe measurements should focus on the determination of the abundances of the heavy noble gases since these latter never condense in the envelopes of Uranus and Neptune and are therefore well mixed, even in the top layers at the $\sim$1-bar level. The trapping properties of noble gases in various icy materials are fairly well known, thanks to laboratory experiments. For example, heavy noble gases are efficiently adsorbed on amorphous ice \citep{Ba07} while Ar is a poor clathrate former because of its small atomic size \citep{Sl08}. The resulting Xe/Ar ratio that desorbed from amorphous ice should then be roughly protosolar in the envelopes of ice giants. In contrast, the same ratio released from clathrates is expected to be supersolar in these envelopes. Also, if the envelopes of the ice giants formed from vapors released at various snowlines instead of a mixture of gas and solids, the abundances of heavy noble gases should be subsolar in their atmospheres. This trend severely departs from the supersolar abundances predicted in the cases corresponding to full clathration or condensation of volatiles in the giant planets' feeding zones. In other words, because noble gases are highly sensitive to the considered mechanism of volatiles delivery, resulting in relative abundances that significantly depend on the delivery process, they should be considered as the top priority of the measurements to be made by an ice giant entry probe.

\begin{acknowledgements}
O.M. and T. C. acknowledge support from CNES. J.I.L. was supported by the Juno project. K.E.M. acknowledges support from from the Rosetta project through JPL subcontract 1585002 and by NASA RDAP grant 80NSSC19K1306. We thank the two anonymous reviewers for their useful comments.
\end{acknowledgements}

\bibliographystyle{spr-chicago}      
\bibliography{example}   
\nocite{*}


\end{document}